\documentclass[a4paper,12pt,final]{article}
\usepackage{epsfig}

\def\beq{\begin{equation}}
\def\eeq{\end{equation}}
\def\p{\partial}
\font\mybb=msbm10 at 10pt
\def\bb#1{\hbox{\mybb#1}}

\topmargin
-1.5cm
\textwidth
15.5cm
\textheight
24cm
\oddsidemargin
0.7cm
\evensidemargin
1.2cm

\begin{document}


\begin{titlepage}
\vfill
\begin{flushright}
CERN-TH/2002-158\\
hep-th/0207178\\
\today\\
\end{flushright}

\vfill
\begin{center}
\baselineskip=16pt
{\Large\bf Dilaton tadpoles and\\
D-brane interactions in compact spaces}
\vskip 0.3cm
{\sl 
\vskip 10.mm
{\bf ~Ra\'ul Rabad\'an
\footnote{Raul.Rabadan@cern.ch},~Frederic Zamora
\footnote{Frederic.Zamora@cern.ch}}\\
\vskip 1cm
{\it  Theory Division, CERN\\
CH-1211 Gen\`eve 23, Switzerland}\\
\vskip.5cm
}
\end{center}
\vfill
\par
\begin{abstract}

We analyse some physical consequences when supersymmetry is broken
by a set of D-branes and/or orientifold planes in Type II string
theories. Generically, there are global dilaton tadpoles at the disk
level when the transverse space is compact. By taking
the toy model of a set of electric charges in a compact
space, we discuss two different effects appearing when global
tadpoles are not cancelled. On the compact directions a constant term
appears that allows to solve the equations of motion. 
On the non-compact
directions Poincar\'e invariance is broken. We analyse some examples
where the Poincar\'e invariance is broken along the time direction
(cosmological models). 
After that, we discuss how to obtain a finite interaction 
among D-branes and orientifold planes in the compact space 
at the supergravity level.

\end{abstract}

\end{titlepage}

\setcounter{page}{1}

\tableofcontents

\section{Introduction}

In recent years, some models with supersymmetry 
broken by a set of D-brane and orientifold planes in Type II string
theories have been considered \cite{s99,susybreaking}. 
When these objects are located 
on a compact manifold, one should take into account that
the Ramond-Ramond tadpoles are cancelled \cite{pc88}. 
On the contrary the NS-NS tadpoles can remain uncacelled
\footnote{There is the possibility of breaking supersymmetry and avoid
the disk tadpoles as in \cite{notadpoles}.}. 
The presence of these tadpoles has a series of physical consequences:
redefinition of the background in such a way 
that Poincar\'e invariance is broken
\cite{fs86}, effective potentials for some moduli \cite{potential},
divergences in higher string amplitudes, etc. Some
supergravity solutions in the presence of these global tadpoles 
can be found in \cite{dm00,bf00,c01}.
                                                                                           
In a pair of papers, Fischler and Susskind  \cite{fs86} showed how                  
to deal with these NS-NS tadpoles and the divergences associated with                     
them. In the first paper, they showed how the divergences can be                       
eliminated by introducing a dilaton condensate. This condensate acts like                 
a background that is a solution to the equations of motion with the                        
inclusion of the tadpole terms. In the second paper, it was shown how to                    
deal with small divergent handles (in our case they will be disks) and how they                 
can be cancelled by shifting to a suitable background \footnote{See,                   
for more detailed explanation, \cite{po98,gr01}.}.                                          

In this paper we intend to re-visit 
some of the physics that arises from a system with
global charges. As a guiding illustrative example,
we take the toy model of a system of electric charges
located at some points in a compact manifold but expanded in
some additional non-compact directions. 
We will see that due to the presence of these global
charges, a tadpole term is generated, having two important physical
consequences. On the one hand, in order to find a solution to the
electrostatic potential, Poincar\'e invariance should be broken. The
term which breaks Poincar\'e invariance is proportional to the sum of
the charges divided by the volume of the compact manifold. When the global
charge vanishes or when we take the decompactification limit,
that term goes to zero, allowing Poincar\'e invariant solutions.

The other effect concerns the compact manifold. The same term that
breaks Poincar\'e invariance appears like a uniform neutralising
background. This is similar to the jellium model in condense matter
physics. There, the ions in a solid are replaced by a rigid uniform
background of positive charge while valence electrons neutralise this
background \footnote{For an introduction to the jellium model
approximation  see, for instance, \cite{z88}.}. In our case, the
jellium term allows to find solutions to the Poisson equation in a
compact space when the total charge is not vanishing. This term changes
the usual behaviour of the propagator in the compact space. The shape
of this propagator is, at short distance with respect to the
compactification volume, similar the the propagator in non-compact
space. But at large distances, the corrections from the jellium term
become important. 
From this redefined propagator we can derive 
the interactions in the compact space. 
As expected, the result coincide with
other methods to define it, like a suitable regularization of 
the potential created by an infinite periodic array of 
images or by computing the energy of the system.

From the electrostatic analogy we pass to the D-brane
case. Now the equations we would like to solve are the Einstein equations
with the dilaton and Ramond-Ramond fields
turned on, with the D-branes being the sources
for these fields. These equations are considerably more complicated
(highly non-linear) than in the simplified electrostatic case but, as we will
see, the physical behaviour is similar. By
integrating the dilaton equation on the compact space, we get that if the
transverse space to the D-branes and/or O-planes is compact and
the sum of the 'dilatonic charges' does not vanish, there is no
solution for the dilaton field which preserves Poincar\'e invariance
on the brane directions. 
If we suppose that the metric has a bi-warped form, {\it i.e.:} that the
dependence on the non-compact coordinates in the compact space comes
by a global factor (and similarly in the compact ones), one can deduce
the jellium term and a dependence between the warped factors and the
dilaton field. Similar conditions have been obtained from the Einstein
equations under the name of Brane Sum Rules \cite{bsr}. In general,
one can obtain an infinite number of these consistency conditions by
taking linear combinations of them (as explained in appendix A).

Non-vanishing tadpole solutions in supergravity have
been analysed previously in \cite{dm00,bf00}. We revisit the
cosmological solutions of Dudas and Mourad \cite{dm00} for the Sugimoto
model \cite{s99} and construct some others by taking T-duality
transformations. One of the main features of these solutions is the
presence of a space-like singularity. At larger times the string
coupling goes to zero and the dilaton term becomes irrelevant (the
disk is a higher order term than the sphere).

From the electrostatic analogy we can also understand the D-brane
interaction in compact space. The naive cylinder diagram is divergent
due to the tadpole term. The divergence is coming from the sum over
windings in the open string picture, {\it i.e.:} the numbers of the images in
the covering space grows faster than the decay with the distance,
similarly to what is happening in the Olbers paradox. One should
correctly define the propagator in the compact space. A jellium type
of term appears in the definition, which allows to find solutions to
the propagator in the compact space. The presence of the jellium term
in the momentum space has the interpretation of the absence of 
a propagating zero mode, making the amplitude finite and
coinciding with the one defined by the sum over images 
when it is correctly regularised.

\section{The effect of Tadpoles in electrostatics}

To understand the meaning of the different tadpoles due to the
D-branes we will develop a toy model: electrostatics in compact
spaces without boundary. As we will see, although it is a simplified model,
it captures most of the physical results we want to point out for the case 
of D-brane systems.

The basic equation we need to solve is the
Poisson equation with sources localised at some points on the manifold
${\cal M}$:
\begin{equation} \label{Poisson}
\Delta_y \phi = \sum q_i \delta(y - y_i)\,.
\end{equation}
Obviously, this equation has solutions only if $\sum q_i = 0$. We will call
this condition the tadpole cancellation condition. This condition comes
from the integration of the above equation on the compact manifold.
Our case is in some way analogous: the tensions of all the branes 
do not add to zero. However in that case
these charged objects are extended 
in some non-compact extra dimensions.
As we will see shortly, is this fact which allows to 
find solutions to the Poisson equation, 
even if there is a global charge in the compact space. 

Since we present the electrostatic analogy for illustrative proposes,
we consider that the metric on the transverse and parallel
dimensions do not mix. 
Then the Laplacian on the whole space can be
split into a compact and parallel space dependence: $\Delta = \Delta_c +
\Delta_{p}$. Let us also consider that the potential $\phi$ can be decomposed
into a sum of a compact dependent part and a parallel dependent part,
$\phi = \phi_c + \phi_{p}$. The Poisson equation can then be written
as
\begin{equation} 
\label{delta}
\Delta \phi = 
\Delta_c \phi_c + \Delta_{p} \phi_{p} = \sum q_i \delta(y - y_i)\,.
\end{equation}
By integrating this equation on the compact space, one finds that
\begin{equation} \label{PoissonNC}
V_{\cal M} \Delta_p \phi_p = \sum q_i\,,
\end{equation}
where $V_{\cal M}$ is the volume of the compact manifold. The equation
(\ref{PoissonNC}) reflects the existence of the tadpole. In the
dimensional reduced theory it appears as a term in the effective
action:
\begin{equation}
S = \int \frac{1}{2}(g^{\mu \nu} \partial_{\mu}\phi \partial_{\nu}\phi
+ \frac{\sum q_i}{V_{\cal M}} \phi)\,.
\end{equation}
It signals that the background should be re-defined and that there is
no solution in the parallel dimensions which satisfies Poincar\'e
invariance.

Let us turn back to the compact space. By using the equation
(\ref{PoissonNC}) into equation (\ref{delta}),
one finds that the equation to solve is not
the Poisson equation in the compact space, which has no solution, but
the modified one:
\begin{equation} \label{PoissonC}
\Delta_c \phi_c = \sum q_i \delta(y - y_i) - \frac{\sum q_i}{V_{\cal
M}}\,.
\end{equation}

So, in the compact space, the tadpole induces a term in the Poisson
equation that allows to find solutions. It can be interpreted as a
constant neutralising background. We will call this term a jellium
term, borrowing the name from solid state physics. 

The interaction between the charges can be obtained by considering the
propagator in the compact space. This propagator satisfies, as we will see
in section 5, the equation
\begin{equation} \label{Propagator}
\Delta_c G(y)  = \delta(y) - \frac{1}{V_{\cal M}}\,,
\end{equation}
where in addition to the usual delta function,
one has to introduce a jellium-type of term, $- \frac{1}{V_{\cal M}}$, which
allows the equation (\ref{Propagator}) to have solution.

As we know, the interaction between the sources can be obtained from the
propagator,
\beq 
{\cal A} = \sum_{ij} q_i G(x_i,x_j) q_j\,. 
\eeq
But there is another way of computing the interaction between the charges
that gives the same result. If we just try, as in the non-compact
case, to consider the potential created by the other branes and to
introduce a probe, the equation without jellium term has no
solution. However, what one could do is to consider the solution of
eq. (\ref{Poisson}) for a system of charges that satisfy the tadpoles
and obtain the energy of the configuration:
\begin{equation} 
E = \int_{{\cal M}} {\nabla \phi}^2 = - \sum_i q_i \phi(y_i)\,.
\end{equation}
The second expression had been obtained by using the Poisson equation
and the Stokes theorem. The energy will depend on the positions of the
charges.

In the case where the tadpoles do not vanish, one can, by solving the
equation (\ref{PoissonC}), obtain an electrostatic potential that
integrated gives the energy of the system:
\begin{equation} \label{energy}
E = \int_{{\cal M}} {\nabla \phi}^2 = - \sum_i q_i \phi(y_i) +
\frac{\sum q_i}{V_{\cal M}} \int_{{\cal M}} \phi\,.
\end{equation}

On the other hand, the electrostatic potential is given by the 
sum of the propagator multiplied by the charges,
\beq 
\label{Potpro} \phi_c = \sum_i q_i G(y,x_i)\,. 
\eeq
It is easy to see that the energy of the system reproduces the
correct interaction behaviour of the charges. Just substituting
(\ref{Potpro}) into (\ref{energy}) one obtains:
\begin{equation} 
E = - \sum_{ij} q_i G(x_i,x_j) q_j + \frac{(\sum q_i)^2}{V_{\cal M}}
\int_{{\cal M}} G(y)\,.
\end{equation}
It means that the energy of the system reproduces the interaction
plus a constant that depends on the global tadpoles.
Notice that if we take the decompactification limit the tadpole terms
vanish. So it suggests that there is a continuous transformation from
the compact to the non-compact cases, where the tadpole is no present
(flux can escape to infinity).

\subsection{An example: charges in a circle}

Let us consider a system of two charges, $q_1$ and $q_2$, at the
points $y_1$ and $y_2$ in a circle of length $L$. Tadpoles are
satisfied if the sum of these charges is zero.

\begin{figure}
\centering \epsfxsize=6in \hspace*{0in}\vspace*{.2in}
\epsffile{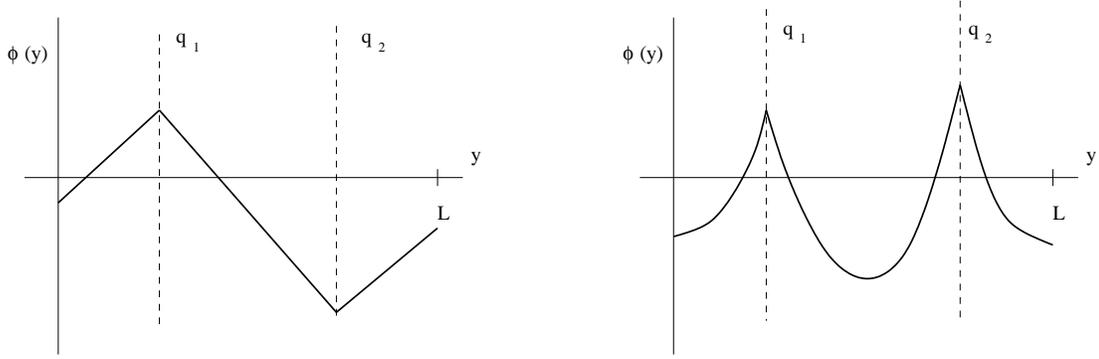}
\caption{\small On the left hand side we have represented the 
electrostatic potential of two opposite charges on a circle. 
As tadpole conditions are satisfied one gets linear dependence as expected. 
On the right side the tadpoles are not cancelled and 
one finds a quadratic dependence on the distance proportional to the tadpole.} 
\label{potential} 
\end{figure}

If, naively, one tries to find by the method of the images
what is the potential created by 
these charges, one finds
\begin{equation}
V(y) = q_1\sum_{n \in \bb{Z}} |y-y_1 + n L| + 
q_2 \sum_{n \in \bb{Z}} |y-y_2 + n L|\,.
\end{equation}
which diverges unless $q_1+q_2 =0$. 
On the other hand, by solving
\begin{equation} \label{PoissonC2}
\Delta_c \phi_c = q_1 \delta(y - y_1) + q_2 \delta(y - y_2) -
\frac{q_1 + q_2}{L}
\end{equation}
and imposing periodicity and continuity, one finds:
\begin{itemize}
\item $0 \leq  y \leq y_1$: $\phi_c(y) = \frac{y}{L}(q_1 y_1 + q_2 y_2 
-\frac{L}{2}(q_1+ q_2)) - \frac{(q_1 + q_2) y^2}{2L} + 
q_1 y_1 +q_2 y_2 +\phi_0$\,,

\item $y_1 \leq  y \leq y_2$: $\phi_c(y) = \frac{y}{L}(q_1 y_1 + q_2 y_2 +
\frac{L}{2}( q_1 - q_2)) - \frac{(q_1 + q_2) y^2}{2L} + q_2 y_2 +\phi_0$\,,

\item  $y_2 \leq  y \leq L$: $\phi_c(y) = \frac{y}{L}(q_1 y_1 + q_2 y_2 +
\frac{L}{2}(q_1 + q_2)) - \frac{(q_1 + q_2) y^2}{2L} + \phi_0$\,.
\end{itemize}
 
The potential is represented in figure \ref{potential}. In the case
where the tadpoles are cancelled there is no quadratic term in the
potential and the electrostatic potential is described by straight lines. 
In the points where there is a charge the first derivative
jumps. It also happens in the case where the tadpoles are not
cancelled. The difference is that there are parabolic segments
instead of lines, as seen in figure \ref{potential}.

\begin{figure}
\centering \epsfxsize=4in \hspace*{0in}\vspace*{.2in}
\epsffile{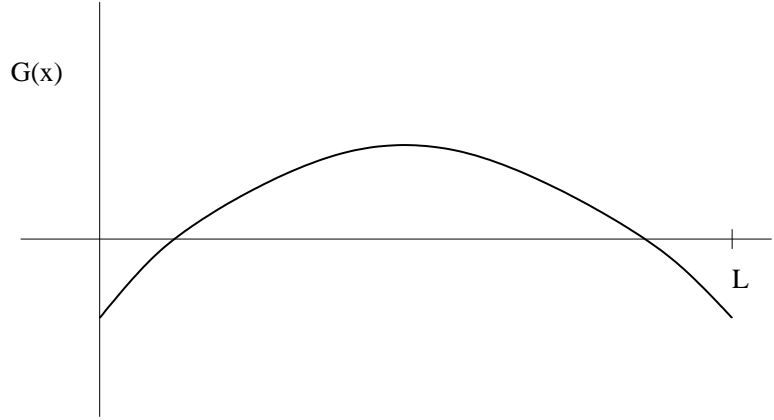}
\caption{\small 
Propagator on a circle. At short distances the behaviour is like in
non-compact space. At large distance should be modified to take into
account that our space is compact.}
\label{propagatordibu} 
\end{figure}

The propagator can also be obtained by soving the equation:
\begin{equation} \label{propa}
\Delta_c G(y) = \delta(y) - \frac{1}{L}\,,
\end{equation}
where we have put the source at the origin. The solution to the above
equation can be easily obtained:
\beq 
\label{sol} G(y) = \frac{|y|}{2} - \frac{y^2}{2L} + C \,, 
\quad\quad |y| \leq L\,.
\eeq
Notice that for large volume the propagator reproduces the expected
behaviour for non compact space, {\it i.e.:} the linear behaviour in our one
dimensional case. At large distances, comparable to the
compactification size, the propagator is modified to be periodic. See
figure \ref{propagatordibu}.

The energy of the system can be computed by using (\ref{energy}):
\begin{equation}\label{E}
E(y_1,y_2) = - q_1 q_2 \left(|y_1-y_2| -\frac{(y_1-y_2)^2}{L}\right) 
+ (q_1+q_2)L \left(C+\frac{L}{12}\right)\,,
\end{equation}
Notice that there is a linear dependence as
expected from a Laplacian type behaviour. Because the space is compact
there is a quadratic dependence (also in the absence of tadpoles). The
tadpoles do not affect the interaction but only give a vacuum energy.

One might wonder if the divergence
appearing by the introduction of the infinite array of images 
can be conveniently regularized and then substracted,
in such a way that the resulting finite potential reproduces
the physical potential and propagator obtained by the
previous methods. The non-compact propagator 
for an infinite periodic array of charged images is
\begin{equation}
G_{\rm div}(y) = \sum_{n \in \bb{Z}} |y + n L| = |y| + \sum_{n \in \bb{N}}|y + n L|
+ \sum_{n \in \bb{N}}|y - n L|\,.
\label{solzero}
\end{equation}
By using (regulating and discarding the divergent part
\cite{po98})
\begin{equation}
\sum_{n > 0} (n - a) = \frac{1}{24} - \frac{1}{8}(2a -1)^2\,,
\end{equation}
one obtains
\begin{equation}
G_{\rm reg}(y) = |y| - \frac{y^2}{L} - \frac{L}{12}\,,
\label{solreg}
\end{equation}
which reproduces the interaction obtained by solving the 
equation (\ref{propa}) with $C=-L/12$.

It is curious to observe that if we express 
the propagator (\ref{solreg}) in Fourier modes, we get 
\beq
G_{\rm reg}(y) = - \frac{L}{4\pi^2}\sum_{n\not=0}
\frac{e^{2\pi i n {y\over L}}}{n^2}\,,
\eeq
{\it i.e.:} the zero mode has been completelly substracted.
Therefore, its inclusion is what made the propagator (\ref{solzero}) 
divergent.

Summarising: there are two different effects of tadpoles in compact
manifolds. From the point of the reduced effective action 
on the non-compact space, they modify the action, meaning that the background 
solutions are redefined. 
From the point of view of the compact dimensions the
tadpole appears as a necessary extra term in the equations of motion 
on the compact space.

\section{Interpretation of NS-NS tadpoles in superstring theories}

Once we have understood the electrostatic case, let us take a system
of branes, antibranes and orientifold planes located at different
points in a compact space.
We will pay attention to the dilaton equation and the consistency
conditions derived from there. The idea is to integrate the equation in
the compact space and, like in the electrostatic case, obtain some
consistency conditions. The physical consequences are analogous to
the electrostatic case: dilaton tadpoles break Poincar\'e invariance
along the directions parallel to the D-branes. The jellium term has
also an analogy in these systems.

Similar consistency conditions can be obtained by considering the
Einstein equations. These ideas have been applied in brane world
scenarios under the name of brane sum rules \cite{bsr}. Even more, one
can consider linear combinations of these consistency conditions as
explained in the appendix.

\subsection{Dilaton equation}

In the Einstein frame, the equation for the dilaton coupled to some
sources (D-branes, orientifold planes,...) has the form:
\begin{equation}
\partial_M\left(\sqrt{-G} G^{MN} \partial_N \phi\right) 
+ \frac{3-p}{4}
(e^{{(p-3)\over 4}\phi}\sqrt{-{\hat g}} \sum_i q_i \delta^{9-p}(y-y_i) 
- e^{{(3-p)\over 2}\phi} \sqrt{-{g}} \frac{|F_{p+2}|^2}{(p+2)!})
=0 \,,
\end{equation}
where $G_{MN}$ is the metric in ten dimensions,
${\hat g}_{\mu \nu}$ the induced metric on the $p$-brane 
and $q_i\sim g_s l_s^{7-p}$ are its tensions.
Finally, $y_i$ are the positions where the different objects are located.

Let us consider the case of a ten dimensional manifold with a topology
$R^{p+1} \times {\cal M}$ where ${\cal M}$ is a compact manifold of
dimension $9-p$. We consider Dp-branes expanding the non-compact
coordinates and located at some points in the compact part. Let us
take the metric
\begin{equation}
G_{MN} = g^{nc}_{\mu \nu}(x,y) dx^{\mu} dx^{\nu} + g^c_{ij}(x,y)
dy^{i} dy^{j}\,.
\end{equation}
With this ansatz for the metric and integrating the equation for the
dilaton we obtain the following consistency condition:
\begin{equation}
\int_{\cal M} \partial_{\mu} (\sqrt{-G} G^{\mu \nu} \partial_{\nu}
\phi)  = \frac{p-3}{4} \sum_i q_i \sqrt{-g^{nc}(x,y_i)} e^{{(p-3)\over 4} \phi(x,y_i)}
+ \frac{3-p}{4}\int_{\cal M} e^{{(3-p)\over 2}\phi} \sqrt{-{g}} \frac{|F_{p+2}|^2}{(p+2)!} \,.
\end{equation}

An immediate physical consequence is that if 
the right hand side of the equation is non-vanishing,
one has to require that $\partial_{\mu} \phi \neq 0$, 
{\it i.e.}: the dilaton field has to break the
Poincar\'e invariance along the directions of the branes.

\subsection{(Bi)warped  metric}

Let us take a less general metric with warped factors in both the
non-compact and compact coordinates,
\begin{equation}
\label{bi_warped}
G_{MN}(x,y) = e^{\Omega(y)} g_{\mu \nu}(x) + e^{B(x)}h_{i j}(y)\,.
\end{equation}
Notice that this case includes the usual warped cases, like the
supersymmetric solutions, where $B(x)=0$ and $g_{\mu\nu}(x) =\eta_{\mu\nu}$.

We can consider, like in the electrostatic
case, that the dilaton can be decomposed in a compact and a
non-compact dependence: $\phi(x,y) = \varphi(x) + \phi(y)$. The action
for the dilaton in the Einstein frame is of the form:
\begin{equation}
S = -\frac{1}{4 \kappa^2} \int \sqrt{-G} \left(e^{-\Omega}|\partial
\varphi(x)|^2_g + e^{-B}|\partial \phi(y)|^2_h \right)\,.
\end{equation}

\vspace{0.3cm}

With this ansatz the dilaton equation can be decomposed into a set of
different equations:
\begin{enumerate}
\item 
There is a linear relation between the warped factor on the
compact coordinates and the  $\varphi(x)$ of the form:
\begin{equation} 
\label{guan}
2 (7-p) \p_\mu B(x) = (p-3) \p_\mu\varphi(x)\,.
\label{dilaton_x}
\end{equation}
This relation comes from the factorisation of the $x$ and $y$
dependence of the dilaton equation.

\item 
A similar equation can be obtained from the previous one and the
Einstein equations for mixed indices (see Appendix B). This equation relates
the warped factor on the non-compact coordinates and the  $\phi(y)$.
Assuming $\p_\mu B\not=0$ \footnote{Below there is a discussion  
for $p=3$, where $\p_\mu B =0$.}, we have
\begin{equation}
\label{dilaton_y}
2 (3-p) \p_i \Omega(y) = (p-7) \p_i \phi(y)\,.
\end{equation}

\item 
By integrating on the compact manifold, we can obtain the dilaton equation 
in the non-compact coordinates:
\begin{equation}
V_{\cal M} \partial_{\mu}(e^{B(x)} g^{\mu \nu} \partial_{\nu}\varphi(x)) 
= \frac{(p-3)}{4} 
(T - E) \label{non-compact_B}
\end{equation}
where $T$ and $V_{\cal M}$ are the analogues of the tadpole term and volume term
in the electrostatic case:
\begin{equation}
T = \sum_i q_i e^{-\frac{8}{p-7}\Omega(y_i)}
\end{equation}
and
\begin{equation}
V_{\cal M} = \int_{\cal M} \sqrt{g_c} e^{\frac{p-1}{2}\Omega(y)}\,.
\end{equation}

The $E$ term is due to the coupling of the RR fields to the dilaton in the Einstein frame:
\begin{equation}
E = \int_{\cal M} \sqrt{h}  e^{- \frac{p-3}{2} \phi(y) - \frac{p+1}{2} \Omega(y)} 
|F_{p+2}|^2
\end{equation}
We will discuss bellow the physical interpretation of these equations.

\item 
The dilaton equation in the compact manifold is
\begin{eqnarray}
\partial_i(\sqrt{h} e^{\frac{p+1}{2}\Omega} h^{ij} \partial_j\phi(y)) & = &\frac{p-3}{4} ( e^{\frac{p-3}{4}\phi + \frac{p+1}{2}\Omega} \sum_i q_i \delta(y-y_i) - e^{\frac{3-p}{2}\phi - \frac{p+1}{2}\Omega} \sqrt{h} 
|F_{p+2}|^2
) \nonumber \\
 & & - \frac{p-3}{4} e^{\frac{p-1}{2}\Omega} \sqrt{h} \frac{T -E}{V_{\cal M}}
\label{dilaton_compact}
\end{eqnarray}
\end{enumerate}

\subsection{RR Field Equation}

We are working up to the disk level in the 
string coupling. At this order, the branes do not
interact among them, since the cylinder contribution is 
at higher order. In this static approximation, we can take
the RR $p+1$ form potential, in an appropriate gauge, to be
\beq
C_{p+1}= C(y)dx^0\cdots dx^p\,.
\eeq

Then, the equation of motion for the RR field is
\beq
\p_i\left(e^{-(\frac{p+1}{2})\Omega + \frac{3-p}{2} (\phi + \varphi)}\sqrt{h}h^{ij}\p_j C\right)
= \left(e^{\frac{p-7}{2}B(x)}\sqrt{-g(x)}\right) \sum_n q^{RR}_n \delta^{9-p}(y-y_n)\,,
\label{RR_equation}
\eeq
where $q^{RR}_n$ are the 
RR charges for the D-brane and orientifold planes.

\subsection{Physical implications}

Now we extract some physical implications from the above equations:

\begin{itemize}

 \item 
The volume $V$ of the compact space has a warped factor
dependence $\Omega(y)$. That can be seen directly by reducing the action on
the compact space with the above ansatz. That dependence also appears
in the supersymmetric case when the NS-NS and RR charges are not
cancelled locally.

 \item 
Also the tadpole $T$ has a warped factor dependence. This
dependence comes from the volume and dilaton dependence of the
coupling.
As in the illustrative example in section 2, 
if the volume goes to infinity or
the tadpole vanishes a solution to the equations of motion can be
found where Poincar\'e symmetry is preserved.
When tadpoles are cancelled like in T-dual models to Type I one
can construct the solution by the harmonic function method by taking
into account all the images. The tadpole $T$ in function of this
harmonic function is of the form:
\begin{equation}
T = \sum_i q_i H^{-1}(y_i).
\end{equation}
One can check that $T$ is equal to zero in these case because the
harmonic function has pole at the points where the D-branes are located,
for $p<7$.

 \item For a system of Dp-branes with $3 < p <7$ the above equations
tell us  that:
\begin{itemize}

\item From equation (\ref{dilaton_x}),
the cosmological evolution of the compact coordinates is related to
$\p_\mu\varphi(x)$ with the same sign.
This means that if the compact coordinates are getting
smaller, the string coupling becomes weaker.

\item As in the supersymmetric case, the warped factor $\Omega(y)$ is
proportional to the dilaton dependence on the compact space 
(up to a constant shift) due to the equation (\ref{dilaton_y}).

\end{itemize}

 \item
The case $p=3$ is special. From equation (\ref{dilaton_x}) we 
get $\p_\mu B(x) =0$. Then, for this case, the equation
(\ref{non-compact_B}) tell us that one can choose a constant value of the dilaton. That was expected, since D3-branes and anti D3-branes do not couple to the dilaton.

 \item
For $p<3$, the $x$-derivatives in the equation (\ref{dilaton_x})
have opposite sign. For a cosmological solution with the compact
dimensions becoming larger in time the string coupling decreases.



\item
Contrary to the case of the NS-NS tadpole, the RR tadpole has 
to cancel, as can be seen by integrating the equation 
(\ref{RR_equation}) on the compact manifold
to obtain $\sum_n q^{RR}_n =0$.

\item
 Since the left hand side of the equation (\ref{RR_equation})
is independent of the brane coordinates $x$, 
we have that, at the disk level 
in perturbation theory, the expansion of the universe
on the brane given by the metric $g_{\mu\nu}$ has to be conveniently
compensated by the expansion of the moduli $B(x)$, such that 
\beq
\label{RR_Eq}
e^{- (\frac{p-7}{2})B(x)}\sqrt{-g(x)}= {\rm const}.
\eeq

 \item 
A similar analysis can be carried out by taking the Einstein equations. 
This conditions are known as Brane Sum Rules \cite{bsr}. One can consider
linear combinations of these condition to obtain a continuous set of
consistency conditions as explained in the appendix A.

\end{itemize}

\section{Examples}

In this section we will consider some examples where dilaton tadpoles
are present. In particular, we will be interested in finding the
solutions to the supergravity equations with the disk terms
present. First we review 
the more familiar supersymmetric case (T-dual
configurations to Type I string) where the supersymmetric
solution only depends on an harmonic function
which can be expressed by the method of the images. 
In this case summing over all the images gives a finite result, 
since there are no tadpoles.
Then we will consider the case where NS-NS tadpoles are not cancelled
but there are no transverse directions to the brane,
such that the RR tadpole is cancelled locally. The only
physical consequence will be the lack of Poincar\'e invariant
solutions on the brane. In particular we will 
re-visit the cosmological solution
of the Sugimoto string \cite{s99} found by \cite{dm00}. Then we will
construct some models with lower dimensional branes 
by taking T-dualities of the above solution.

\subsection{Supersymmetric case}

Let us consider as an illustration a T-dual model to Type I theory on
a torus. The Dp-branes are located at some points of a $T^{9-p}$
torus. There are also $2^{9-p}$ orientifold planes with RR charge and
tension equal to $2^{p-5}$ in Dp-brane units. In order to cancel the
RR-charge 16 Dp-branes are needed.  One can cancel the tension and RR
charges locally if $p \geq 5$ by putting $2^{p-5}$ Dp-branes on each
Op-plane.

The solution to the supergravity equations with these sources in a
compact space can be constructed from the harmonic function in the
compact space \footnote{We take an squared torus to symplify notation.}:
\begin{equation}
H(y) = 1 + \sum_i q_i \sum_{\vec{m_i}\in \bb{Z}^{9-p}}
|\vec{y_i} + \vec{m_i} L|^{p-7}\,.
\label{harmonic}
\end{equation}

Notice that this function does not suffer from the tadpole divergence
due to the fact that the sum of all the charges is zero (so the
constant term vanishes). The metric, dilaton and RR field are 
determined by as the harmonic funcion (\ref{harmonic}). There is a
problem close to the orientifold planes because the $H$ change its
sign, and the metric is not defined with a negative harmonic
functions. That is expected to be cured by non-perturbative effects in
the same way as for the $O6$-plane non-perturbative effects change
the Taub-NUT metric with negative charge into the Atiyah-Hitchin
one. These effects are expected to be important close to the
orientifold planes but the solution constructed from the above
harmonic function is expected to be correct far away from the negative
tension objects.

Finally, notice that this solution respects Poincar\'e symmetry 
along the directions parallel to the branes as we expected 
from tadpole cancellation conditions.

\subsection{Sugimoto model}

The first example we consider is the Sugimoto model \cite{s99}. The
idea is, like the orientifold construction of Type I string, to start
from Type IIB and introduce an orientifold plane with opposite charges
(NS-NS and R-R) with respect the Type I orientifold. In order to cancel the R-R
tadpoles one introduces 16 dynamical antiD9-branes (32 if the
orientifold images are taken into account). These antibranes break down
the supersymmetries that were preserved by the
orientifold, such that no supersymmetry is left unbroken. 
From the open string projection one can see that the
orientifold keeps the symmetric representation for the massless gauge
bosons and the antisymmetric one for the gauginos. All together gives a
$USp(32)$ gauge group with fermions transforming in the antisymmetric
representation.

As NS-NS charges are not cancelled, there is a tadpole term in the
effective action. This configuration is the extreme case where the
compact space is just a point, and all the non-trivial behaviour due
to the NS-NS tadpoles is reflected in a breaking of the Poincar\'e
invariance at the disk level. The re-definition
of the background due to these terms affects the closed string
propagator beyond one loop.
But at the level of the disk, there are  already effects of the tadpole term in the ten
dimensional background. They were analysed by Dudas and Mourad
\cite{dm00}. The effective potential is of the form (in Einstein
frame):
\beq 
\int \Lambda e^{-3 \phi/2}\,,
\eeq
where $\Lambda = 32 T_9$.
This potential looks very similar like a quintessence potential, but
the solution has a very different behaviour. We will consider
homogeneous and isotropic spatially flat solutions.
The solution has a space-like singularity and 
the dilaton potential energy and the ten dimensional 
scalar curvature go to zero at the infinite future. The
universe at long times suffers a deceleration. It is easy to
demonstrate that the solution is unique up to
two parameters that determine the time location of the 
(big bang) singularity and the normalization of the string coupling. 

\begin{figure}
\centering \epsfxsize=4in \hspace*{0in}\vspace*{.2in}
\epsffile{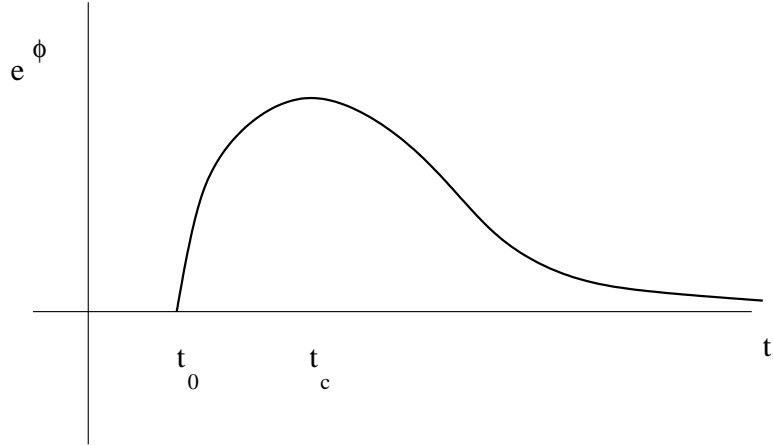}
\caption{\small 
Dilaton behaviour.}
\label{dil} 
\end{figure}

The solution in the Einstein frame has the following form \cite{dm00}:
\begin{eqnarray}
e^{\phi} & = & e^{\phi_0} (\sqrt{\Lambda} t - c_0)^{2/3} e^{\frac{-3
(\sqrt{\Lambda} t - c_0)^2}{4}} \nonumber \\ ds_E^2 & = & -
(\sqrt{\Lambda} t - c_0)^{-1} e^{-3\phi_0/2} e^{\frac{9 (\sqrt{\Lambda} t -
c_0)^2}{8}} dt^2 + (\sqrt{\Lambda} t - c_0)^{1/9}
e^{\frac{(\sqrt{\Lambda} t - c_0)^2}{8}} dx_{||}^2\,,
\label{9metric}
\end{eqnarray}
where  $c_0$ and $\phi_0$ are the constant 
parameters of the solution. The value of the $c_0$ is related to the
position of the singularity by $t_0 = -c_0/\sqrt{\Lambda}$.

The dilaton grows from zero to a maximal value at $t_c = (2/3
-c_0)/\sqrt{\Lambda}$. The value of the dilaton at the maximum is
$e^{\phi_c} = e^{\phi_0} (2/3)^{2/3} e^{-1/3}$. Then it starts to decrease to
zero, as seen in figure \ref{dil}.

\begin{figure}
\centering \epsfxsize=4in \hspace*{0in}\vspace*{.2in} \epsffile{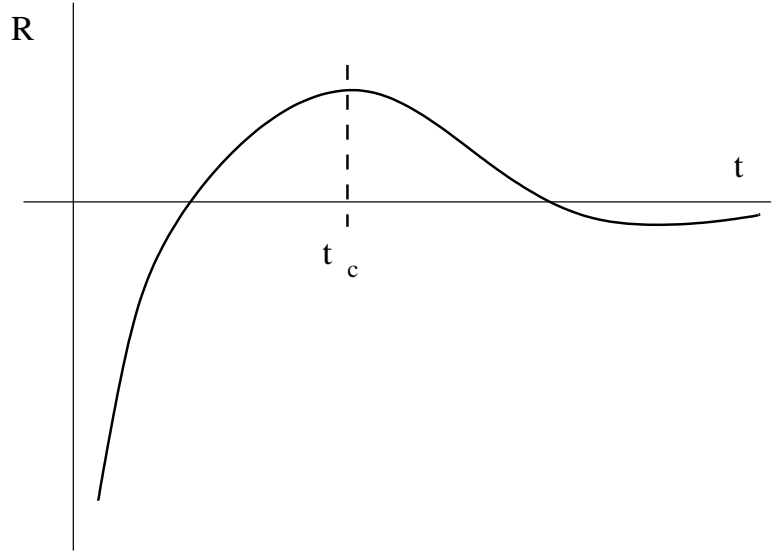}
\caption{\small Ricci scalar for the Dudas-Mourad solution.} 
\label{R} 
\end{figure}

The scalar curvature of the solution when $t_0=\phi_0=0$ is 
(for other values the solution is
physically equivalent; see figure \ref{R}):
\beq R = \frac{-81 t^4 +252 t^2 -16}{72 t e^{9t^2/8}}\,. \eeq

The maximum of the Ricci scalar is at $t_c$, the maximum also of the
dilaton potential. 
Close to the critical point $t_c$, where the dilaton gets its maximum,
the universe is accelerating. The energy is basically concentrated in
the potential of the dilaton that acts as a cosmological constant.
At the singularity, $e^{\phi}$ goes to zero and the scalar curvature
diverges as $(t-t_0)^{-1}$. The dilaton potential goes to zero.

For larger times, the dilaton potential energy goes to zero, the
universe deccelerates. Strings are very weakly coupled and the system
is driven to free strings. The solution is not flat in that case
because there is some energy in the dilaton kinetic term.

The Poincar\'e limit can be obtained when the tadpole $\Lambda$ 
goes to zero. In that case the singularity goes to
infinity, the metric becomes the constant flat metric $\eta_{\mu\nu}$ 
and the dilaton gets a constant value 
$e^{\phi} = e^{\phi_0}(c_0)^{2/3}$ 

\subsection{T-duality}

Since the metric is translationally invariant on the spatial coordinates 
$x_{\parallel}$, we can perform T-duality transformations along these
directions. Consider compactifying one direction: $x^9 \simeq x^9 +1$. 
The T-dual metric to (\ref{9metric}) is, in the Einstein frame:
\begin{eqnarray}
ds_E^2 =&& - (\sqrt{\Lambda} t - c_0)^{-1} c_1^{-3/2} 
e^{\frac{9 (\sqrt{\Lambda} t - c_0)^2}{8}} dt^2 + 
(\sqrt{\Lambda} t - c_0)^{1/9} 
e^{\frac{(\sqrt{\Lambda} t - c_0)^2}{8}} d\hat{x}_{||}^2
\nonumber
\\
&&+ (\sqrt{\Lambda} t - c_0)^{2/3} e^{-{9\over 8}(\sqrt{\Lambda} t - c_0)^2}
dx_9^2
\end{eqnarray}
and the dual dilaton becomes
\beq
e^{\tilde{\phi}}=e^{\tilde{\phi_0}}(\sqrt{\Lambda} t - c_0)^{7/9}
e^{-{5\over 8}(\sqrt{\Lambda} t - c_0)^2}\,.
\eeq

This solution corresponds to an smeared distribution on the
9-direction of orientifold 8-planes and anti-D8-branes. Notice that the
solution has no singularities in the internal space. The solution
where the D8-branes are localised on the top of each of the orientifold
planes was constructed in \cite{bf00}, finding  
singularities in the internal space where also the
dilaton diverges. It would be nice to see how general is the relation
between localised solutions and singularities in the compact manifold.

With only one direction T-dualized,
the scalar curvature and the dilaton profile are very similar to the 
original model. We can proceed performing further T-dualities on 
the extra spacial directions. In the string frame, our original metric
has the expression
\beq
ds_{\rm string}^2 = -a_{\rm string}^2(t) dt^2 
+b_{\rm string}^2(t) dx_{\parallel}^2\,,
\eeq
with
\begin{eqnarray}
a_{\rm string}^2(t) &=& e^{-\phi_0} (\sqrt{\Lambda} t -c_0)^{-2/3}
e^{{3\over 4}(\sqrt{\Lambda} t -c_0)^2}\,,
\\
b_{\rm string}^2(t) &=&e^{-\phi_0/2} (\sqrt{\Lambda} t -c_0)^{4/9}
e^{-(\sqrt{\Lambda} t -c_0)^2/4}\,.
\end{eqnarray}
We can perform $n=9-p$ T-dualities on the longitudinal directions 
$\vec{x}_{\parallel}$ to obtain additional
brane transverse directions $\vec{x}_{\perp}$. 
According to Busher's rules, 
the following background is also a solution
of the equations of motion, corresponding
to having $p=9-n$ D-branes and orientifold planes
dislocalized in the $n$ compactified directions $\vec{x}_\perp$:
\begin{eqnarray}
ds_{\rm string}^2 &=& -a_{\rm string}^2(t) dt^2 
+b_{\rm string}^2(t) dx_{\parallel}^2 +b_{\rm string}^{-2}(t) dx_{\perp}^2\,,
\\
\nonumber
e^{\tilde{\phi}} &=& e^{\phi(t)} b_{\rm string}^{-n}(t)
\\
&=& e^{\frac{p-5}{4}\phi_0} (\sqrt{\Lambda} t -c_0)^2)^{\frac{2}{9}(p-6)}
e^{\frac{3-p}{8}(\sqrt{\Lambda} t -c_0)^2}\,.
\end{eqnarray}

Notice that for $3<p<6$, the string coupling diverges at $t=t_0$
and decays exponentially for $t>t_0$. At $p=3$, the coupling
follows only a power law dependence in $t$. This
case corresponds to a system of dislocalized orientifold 3-planes and
anti-D3 branes along the 
six transverse directions. In this way, RR flux is locally cancelled and
the transverse translational invariance is preserved.
Since in the Einstein
frame the dilaton does not couple to 3-branes,
its decreasing is much slower 
than the scalar curvature's, which in the Einstein frame goes as
\beq
R(t) =\frac{(2-9 (\sqrt{\Lambda} t -c_0)^2)}
{(\sqrt{\Lambda} t -c_0)^{5\over 3}e^{\frac{3}{4}(\sqrt{\Lambda} t -c_0)^2}}\,.
\eeq

\begin{figure}
\centering \epsfxsize=6in \hspace*{0in}\vspace*{.2in}
\epsffile{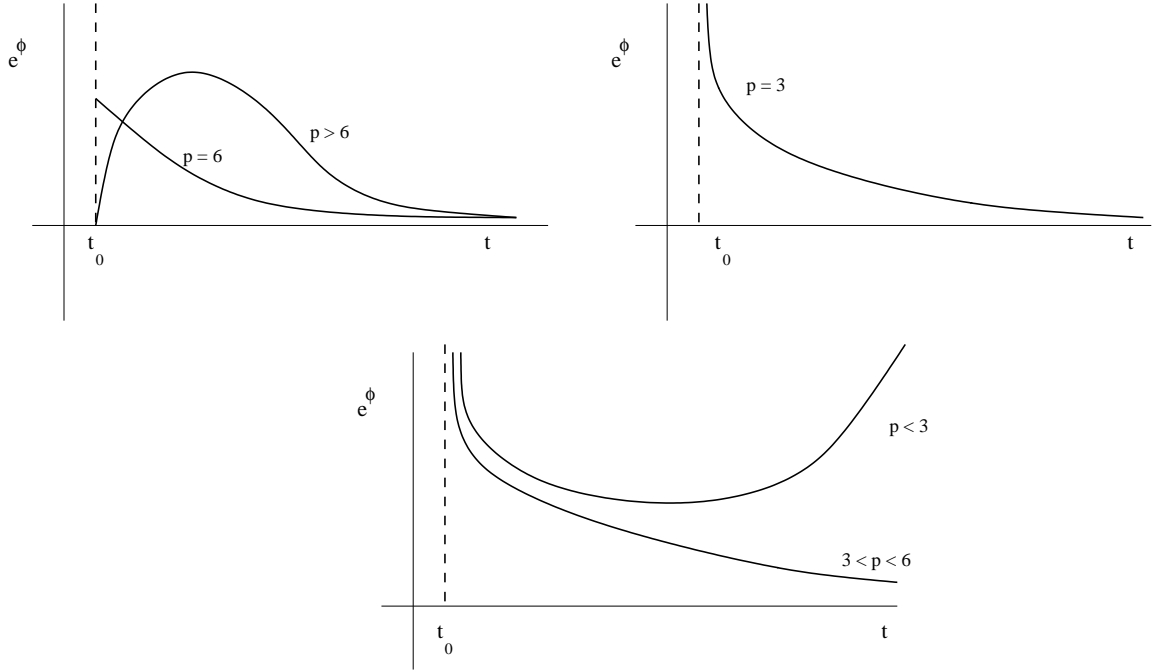}
\caption{\small 
Dilaton behaviour for different solutions depending on the dimension of the brane.}
\label{dil2} 
\end{figure}

When $p > 6$ we have the same qualitative behaviour as in the Sugimoto
solution in ten dimensions: at large times and close to the
singularity the solution goes to weak coupling. The case $p = 6$ is
interesting as the solution near the singulaty has a non-vanishing
coupling but at large distances keeps the exponential decay. For $p <
3$ the dilaton has a divergent behaviour at the singularity and at
the infinite future.

Another interesting point is that the solutions we describe here do
not satisfied the relation (\ref{guan}). That is because for deriving
that relation one assumes that the sources are localised, or at least
have some dependence, on the compact manifold. However 
for these T-dual solutions 
that is no longer true, and one can find solutions where
there is no dependence on the compact coordinates. For localised or
semi-localised solutions, like the ones in \cite{bf00}, the relations
hold.

\section{D-brane interactions}

\subsection{Introduction}

Let us consider some D-branes expanding some non-compact space
coordinates and being localised at some points on a compact
manifold. Ramond-Ramond tadpoles impose strong constraints on the
allowed configurations: without orientifold planes, the
number of branes should be equal to the number of antibranes. As
brane-antibrane forces are attractive and the the brane-antibrane pair
can anihilate each other, one expects the supersymmetric vacuum to be the
final configuration.  This is not the case in the presence of
orientifold planes (for example, for T-dual configurations of the Type
I superstring),  where a net RR charge for the D-branes is necessary
to cancel the negative RR charge of the orientifold planes.

If the configuration is supersymmetric, the D-brane interactions are
absent; this is easy to check by the tree level exchange of closed
strings. Things can be slightly different when the disk terms are
present, since they are the dominant terms in perturbation theory. A very
clear example is the case of a brane and antibrane, at just opposite
points on a compact space. The situation is unstable at one loop due
to the attractive force between the two branes. However the disk terms
make  the compact space to expand, taking both branes far appart,
such that non-supersymmetric sectors become very massive
and decouple.

Let us for the moment forget about the disk terms and pay
attention to the one loop amplitudes. From the open string point of
view supersymmetry allows a cancellation between bosons and
fermions. That means that the field which represents the distance
between the branes has a flat direction.
But when supersymmetry is not present, there is
a one loop potential for the scalar fields, which represent the
distances between the two branes. From the closed string point of view
it means an interaction between the D-branes at tree level.

Let us first review the interaction between a Dp-brane  and and
anti-Dp-brane in flat space when the transverse directions  are
non-compact. By going to the one loop open string channel  that
interaction is just the vacuum energy of the system \cite{bs95,po98},
\begin{equation}
V(y_i,\theta) = - V_{p+1} (i)^p \int_0^\infty \frac{dt}{t}(8 \pi^2
\alpha'\,t)^{-\frac{p+1}{2}}  e^{-(\frac{y^2}{2 \pi \alpha'}-\pi) t}
\eta^{-12}(it) \theta^4_{11}(it/2,it)\,,
\end{equation}
where $y$ is the modulus of the vev of the scalar fields 
living on the brane, parametrizing the transverse directions. 
In the closed string
interpretation $y$ is the distance between the two branes.  At long
distances with respect the string length, 
the amplitude is better reproduced by performing a Poisson
re-summation,
\begin{equation}
V(y_i,\theta) = - V_{p+1} (i)^p \int_0^\infty \frac{dt}{t}(8 \pi^2
\alpha'\,t)^{-\frac{p+1}{2}}  e^{-(\frac{y^2}{2 \pi \alpha'}-\pi) t}
t^4 \eta^{-12}(i/t) \theta^4_{11}(1/2,i/t)\,.
\end{equation}
The main contribution is for $t \rightarrow 0$,
\begin{equation}
V(y_i,\theta) = - V_{p+1} (i)^p \int_0^\infty \frac{dt}{t}(8 \pi^2
\alpha'\,t)^{-\frac{p+1}{2}}  e^{-\frac{y^2}{2 \pi \alpha'}t} 16  t^4
\simeq y^{p-7}\,.
\end{equation}

As expected from the supergravity analysis, at long distances the
interactions between the branes are mediated by massless closed string
fields. The potentials produced by these fields are the potentials
expected from the solutions of the Laplacian operator in the
directions transverse to the branes.
The amplitude is not divergent except at short distances, when the open
string tachyon develops. This divergence can be cured by analytic
continuation. We can obtain a real part (the potential)
and an imaginary part associated to the tachyon that reflects that the
system can decay to another system with lower energy (in this case the
tachyon).

After this short review of the one-loop amplitude 
for non-compact transverse space, 
we compactify $9 - p$ transverse directions to
the D-branes in a torus. In the open string one loop 
amplitude, one should
take into account the contributions from the winding modes,
\begin{equation}
V(y_i,\theta) = - V_{p+1} (i)^p \int_0^\infty \frac{dt}{t}(8 \pi^2
\alpha'\,t)^{-\frac{p+1}{2}}  \sum_{m_i} e^{- {t\over2\pi\alpha'}
\sum_i (y_i + m_i L_i)^2} \frac{\theta^4_{11}(it/2,it)}{e^{\pi t}
\eta^{12}(it)}\,.
\end{equation}

If we naively commute the integral and the sum on the windings we
arrive to the following expression for the potential when the two
D-branes are far apart:
\begin{equation}
V(y) = k \sum_{m_i} [\sum_i (y_i + m_i L_i)^2]^{p-7}
\label{images}
\end{equation}
for some constant $k$. Notice that the transposition of the sum and the
integral allows to interpret our result as the sum over all the
images, as in electrostatics. However, as the number of compact
directions is $9-p$, at a 'distance' in the transverse space $r$, there
are $r^{8-p}$ images, which grows faster with the distance that the
decreasing of the interaction \footnote{That is similar to the Olbers
paradox, where the number of light sources at a given distance is of
the same order as the luminosity, so the total luminosity remains
constant.}. So we have found an IR divergence due to the 
massless closed string modes. 
In the open string picture this divergence is an UV effect 
\footnote{See the appendices of \cite{infla}}.

This divergence was expected, since the system, although is
free of RR tadpoles, suffers from NS-NS tadpoles. These tadpoles
appear at the disk level and indicate that the background should be
redefined \cite{fs86}. But as we have shown in the example of electrostatics,
there is another effect for a tadpole in a compact space: the zero mode
for a massless field has to be excluded from propagating.
It is the inclusion of the zero mode which is causing the divergence in
equation (\ref{images}). In the next subsection, we will derive the correct 
equation for the propagator of massless fields in compact spaces,
such that the exclusion of the zero mode is guaranteed.


\subsection{Propagator in Compact Spaces}

Here we will review how to obtain the propagator using path integral
methods and will verify the volume dependent extra term for its 
differential equation. 

Let us consider a $D$-dimensional compact manifold $\cal{M}$ 
provided with the Euclidean metric $g$. 
It has the finite volume $V=\int dx^D \sqrt{g}$
and no boundaries. Consider the Euclidean action of a massless field
$\phi$ coupled to the source $J$:
\begin{equation}
\label{action}
S(\phi,J) = \int dx^D \sqrt{g} \left({1\over 2}|\p\phi|^2 -\phi J\right)\,.
\end{equation}
In order to compute the generating functional 
\begin{equation}
Z[J] = \int {\cal D}\phi\ e^{-S(\phi,J)}\,,
\end{equation}
we define its path integral measure 
by introducing a basis of ortonormal eigen-functions
$\{\phi_n\}_{n\in \bb{Z}}$, satisfying
\beq
\int dx^D \sqrt{g}\phi_n \phi_m = \delta_{n,m} 
\eeq
and
\beq
\nabla^2\phi_n = \frac{1}{\sqrt g}\p_i(\sqrt{g}\, g^{ij}\p_j\phi_n)
= -\omega_n^2 \phi_n \quad\quad n\in \bb{Z}\,.
\eeq
Notice that $\omega_n \geq 0$.
If $\omega_0 = 0$, then we have a zero mode. 
Normalisation determines that 
$\phi_0= {1\over \sqrt{V}}$. Observe that in the infinity
volume limit the zero mode vanishes.
The existence of this 
zero mode is due to the symmetry $\phi \to \phi + c$ for the 
operator $\nabla^2$. In a more general situation, 
zero-modes always come associated 
to symmetries of the action. 
If a mass term is introduced into the action (\ref{action}),
the new operator in the quadratic action is $\tilde{\nabla}^2
=\nabla^2 - m^2$ and 
its eigen-values simply get shifted by $\tilde{\omega}^2 =
\omega^2 + m^2$.

Using the fact that for any function $\phi(y)$ defined on $\cal{M}$
we have that
\beq
\phi(y) = \sum_n a_n \phi_n(y)\,,
\eeq
we can define the path integral measure by
${\cal D}\phi \equiv \prod_n \sqrt{2\pi}da_n$.
The path integral can now be evaluated: 
\begin{eqnarray}
\nonumber
Z[J] &=& \int \Pi_n (\sqrt{2\pi} da_n)  
\exp{\sum_n(-{1\over 2}\omega_n^2 a_n^2 + a_n J_n)}
\\
\nonumber
&=& \left(\int da_0 e^{a_0 J_0}\right)\prod_{n\not=0}\left( \int_{-\infty}^\infty 
\sqrt{2\pi}da_n e^{-{1\over 2}\omega_n^2 a_n^2 +{J_n^2\over 2\omega_n^2}}\right)
\\
\label{genfunc}
&=&\delta(J_0) ({\rm det}'-\nabla^2)^{-1/2}
\exp\left(-{1\over 2}\int d^D x\ d^D y J(x) G(x-y) J(y)\right)\,,
\end{eqnarray}
where we have shifted the integration variable $a_n \to a_n 
+\omega_n^{-2} J_n$ and introduced the Green function
\beq
\label{Green2}
G(x-y) = - \sum_{n\not=0} \frac{\phi_n(x)\phi_n(y)}{\omega_n^2}\,,
\eeq
which satisfies 
\beq
\label{Green}
\nabla^2 G(x) = \delta^D(x) -{1\over V}\,.
\eeq
One can check that shifting 
$\phi(x) \to \phi(x) -\int d^D y\ G(x-y) J(y)$ into (\ref{action}),
only for the Green function satisfying (\ref{Green}) the 
generating functional (\ref{genfunc}) is reproduced.

For localised charges, the source
does not have a constant component and $J_0 =0$. In this case,
the quadratic action (\ref{action}) does not depend on the zero-mode $a_0$,
The situation is similar to a gauge symmetry, where there is an integral
on a local field which does not appear in the action. As in the local 
symmetry situation, we have to ``gauge-fix'' the translational 
symmetry responsible of the zero-mode.
Following the usual techniques, 
we add the following factor in the path integral:
\beq
{1\over V}\int d^n x\ \int dc\ \delta\left(\phi(x) - c\right) =1\,.
\eeq
Permuting some integrals, such that the one on the translational symmetry,
$\int dc$, is left at the end, we have
\begin{eqnarray}
Z[J]&=&\int dc \int da_0 {1\over V}\int d^D x\ 
\delta\left(\frac{a_0}{\sqrt{V}} - c + \cdots\right)\cdot
\nonumber
\\
\nonumber
&&\cdot \prod_{n\not=0}\left( \int_{-\infty}^\infty 
\sqrt{2\pi}da_n e^{-{1\over 2}\omega_n^2 a_n^2+{J_n^2\over 2\omega_n^2}}\right)
\\
&=&\left(\int dc\right) \sqrt{\frac{V}{{\rm det}'(-\nabla^2)}}
\exp\left(-{1\over 2}\int d^n x\ d^n y J(x) G(x-y) J(y)\right)\,.
\end{eqnarray}

\subsection{Physical consequences in D-brane systems}

We are going to discuss about the physical consequences of the zero
mode in the interaction amplitude between D-branes
located at different points in a compact manifold $\cal{M}$. We know
that there are two ways of understanding this amplitude, as a one loop
amplitude of open strings or as a tree level amplitude of closed
strings. When the sources for the closed strings are 
at distances greater that the string length, 
the amplitude then has the structure:
\begin{equation}
{\cal A} = \sum_{ij} q_i G(x_i,x_j) q_j\,,
\end{equation}
where $q_i$ are the (NS-NS and RR) charges for the D-branes and $G(x,y)$ is the
propagator in compact space of the massless modes mediating the
interaction at large distances.

From the previous discussion we have seen that the propagator in
compact space has a volume dependence. That has several consequences
depending on how we look at this:

\begin{itemize}
\item 
The correct Green equation for a massless propagator in a compact space 
is equation (\ref{Green}).
The volume dependent term is necessary in order to find a solution to the
equation (\ref{Green}).

\item 
We can see that the volume dependent term has the effect of
subtracting the zero mode, since
\beq
\delta^D(x-y) -\frac{1}{V} = \sum_{n\not=0} \phi_n(x)\phi_n(y)\,.
\eeq
Also, form equation (\ref{Green2}) we can see that the zero mode
has been explicitly subtracted from the sum. Its inclusion 
would induce a divergence on the amplitude.

\item 
In toroidal compactifications, if we write the propagator with
theta functions, we can see that the absence of the zero mode is
equivalent to adding a constant counter-term, of the form:
\beq 
G \sim \int_0^{\infty} dx [\Pi_{i=1}^D \theta_3(y_i/L_i,
ix/L_i^2) -1]\,.  
\eeq
It is straightforward to check out that the $-1$ in the previous 
integral is cancelling the contribution of the zero mode.
If we introduce a regulator mass $m$ for the massless closed string state,
one can see that the counter-term in the interaction amplitude 
between the two D-branes is proportional to
\beq 
{\cal A}_{ij} \sim \lim_{m \rightarrow 0} \frac{q_i q_j}{m V}\,.
\eeq
Notice that in the non-compact case the amplitude becomes finite and
there is no need to introduce the counter-term. 

\item 
For the total amplitude, summing over all the D-brane
contributions, the counter-term is of the form:
\beq {\cal A} = \sum_{ij} {\cal A}_{ij} 
\sim \lim_{m \rightarrow 0} \frac{(\sum_i q_i)^2}{m V}\,.
\eeq
Firstly notice that in the non-compact limit the counter-term vanish
and the amplitude is finite. Notice also that the amplitude is finite
if the sum over all the charges is zero, {\it i.e.:} tadpole is cancelled. 
in this case, the open string amplitude gives the correct behaviour.

\end{itemize}

\section{Conclusions}

In this paper we have anayzed two different effects by the NS-NS tadpoles. 
On the one hand, we have seen that the tadpole generically induces solutions 
which are not Poincar\'e invariant on the brane longitudinal non-compact dimensions.
We analyzed cosmological solutions of the Sugimoto model and its T-duals.
As expected, there is an space-like singularity at some time $t_0$.
Later on, the string coupling goes to zero and the space-time becomes flat,
except for $p<3$, where the exponential dilaton grows with time.
Since this solutions where obtained applying T-duality, the $p$-branes 
and orientifold planes are dislocalized on the compact space.
It would be interesting to analyse solutions where the   
D-branes and Orientifolds are located at particular
points in the compact space, as it is done in \cite{bf00}. 

Our approach is perturbative in the string coupling. 
We have seen that, already at the disk level, which is the lowest order
where NS-NS tadpoles appear, there are non-trivial relations 
among the metric on the brane, the warp factor in front of the metric
of the compact space, and the dilaton dependence on the brane coordinates.
These are the equations (\ref{guan}) and (\ref{RR_Eq}) respectively.
We should keep in mind that these relations are derived at the disk level
and for the bi-warped metric ansatz (\ref{bi_warped}).

The same kind of effect happens for the dependence on the compact coordinates.
The NS-NS tadpole produces an extra term in the equations of motion 
of the massless fields, a ``jellium'' type of term. For 
the bi-warped metric ansatz (\ref{bi_warped}), the extra term is 
the last one on the right hand side of equation (\ref{dilaton_compact})
Then, we expect the tadpole to modify also the background field 
profile on the compact space.

The second important consequence of the tadpole that we wanted to
stress concerns the interaction among the charged objects on the 
compact manifold. First we have observed that the periodic interaction
potential for D-branes constructed by the method of images is generically
divergent. A closer analysis shows the zero-mode as the responsable for this 
divergence. Using path integral methods,
we have observed that the proper two-point correlator function 
in a compact space satisfies a ``modified'' Poisson equation, 
the equation (\ref{Green}), where the volume
dependent extra term has the effect of subtracting the zero-mode
contribution form the propagator, dealing with a perfectly finite and consistent
interaction amplitude for D-branes in compact spaces. This new term
is related to the tadpole in the background field equations of motion.
When the two charged
objects are very close with respect the size of the compact space,
the propagator behaves as the propagator in non-compact spaces without
tadpoles. But for distances comparable to the compactification scale, the 
effect of the tadpoles becomes relevant, as we have seen for the illustrative
model of electric charges in a circle.

\bigskip

{\large \bf Acknowledgements}

We have benefited form discussions with
L. \'Alvarez-Gaum\'e, 
R. Emparan, 
J. Garcia-Bellido, 
L. Ib\'a\~nez,
B. Janssen,
F. Marchesano,
A. Uranga, 
M. A. Vazquez-Mozo
and G. Veneziano.

\section*{Appendix A: Total derivative operators in compact spaces}

Let us consider $A(y)$ a function from  compact space to the real
numbers.  For every function $f(A)$ one can define the operator
\begin{eqnarray}
\Delta(f(A)) & = & \frac{1}{\sqrt{g}} \partial_i(\sqrt{g} g^{ij}
\partial_j f(A)) \nonumber \\ & = & f'(A) \Delta(A) + f''(A) (\partial
A)^2\,,
\end{eqnarray}
where $(\partial A)^2 = g^{ij} \partial_i A \partial_j A$ and
$f'(A)=\frac{df(A)}{dA}$.
In particular we can always write the term
\beq \Delta A + g(A)(\partial A)^2 = \frac{1}{f'(A)}(f'(A) \Delta A +
f''(A) (\partial A)^2)\,,  \eeq
where $g = f''/f'$. That allow as to write
\beq \Delta A + g(A)(\partial A)^2 = e^{-\int g} \Delta f\,.  \eeq

Particular cases of interest are:

i) When $g(A)= r$ is a constant,
\beq \Delta A + r (\partial A)^2 = e^{-rA} \Delta(e^{rA})\,.  \eeq

ii) When $g(A)= (n-1)/A$, then
\beq \Delta A + \frac{n-1}{A} (\partial A)^2 = \frac{A^{1-n}}{n}
\Delta(A^n)\,.  \eeq

\subsection*{Linear combinations and consistency conditions}

Now, let us consider a set of equations of the form:
\begin{eqnarray}
\Delta A + r_1 (\partial A)^2 & = & F_1 \nonumber \\ & \ddots &
\nonumber \\ \Delta A + r_N (\partial A)^2 & = & F_N\,,
\end{eqnarray}
where $r_i$ are constants and $F_i$ are general formulae.

One can consider a general linear combination of the equations, that
can be written as
\beq \Delta A + \alpha (\partial A)^2 =  \frac{\sum \lambda_i
F_i}{\sum \lambda_i}\,,
\eeq
where $\alpha = \sum r_i \lambda_i / \sum \lambda_i$. That can be
written by using the results from the previous subsection as
\beq \Delta (e^{\alpha A}) = e^{\alpha A} \frac{\sum F_i
\lambda_i}{\sum \lambda_i}\,.  \eeq

We can integrate the relations on the compact manifold without
boundary and get a set of consistency conditions:
\beq \int_{\cal M}  e^{\alpha A} \sum_i F_i \lambda_i \,.\eeq
There are a $RP^{n-1}$ set of consistency conditions (the linear
combinations up to a global factor).

\section*{Appendix B: Ricci tensor for (bi)warped metrics} \bigskip

We take two warped factors $\Omega(y)$ and $B(x)$. With this ansatz for the
metric the Ricci tensor in ten dimensions is
\begin{eqnarray}
R^{10}_{\mu \nu} & = & R^{p+1}_{\mu \nu} -
\frac{9-p}{2}(\nabla_{\mu}\nabla_{\nu}B + \frac{1}{2} \partial_{\mu}B
\partial_{\nu}B) - g_{\mu \nu} \frac{e^{\Omega-B}}{2} (\Delta \Omega +
\frac{p+1}{2} (\partial \Omega)^2)\,, \nonumber 
\\ 
R^{10}_{\mu i} & = & 2
\partial_{\mu}B \partial_{i}\Omega\,, \nonumber 
\\ R^{10}_{ij} & = &
R^{9-p}_{ij} - \frac{p+1}{2}(\nabla_{i}\nabla_{j}\Omega + \frac{1}{2}
\partial_{i}\Omega \partial_{j}\Omega) - g_{ij} \frac{e^{B-\Omega}}{2} (\Delta B +
\frac{9-p}{2} (\partial B)^2)\,.
\end{eqnarray}
And the Ricci scalar is
\begin{equation}
R^{10} = e^{-\Omega}[R_x - (9-p)(\Delta B + \frac{10-p}{4}(\partial
B)^2)]+ e^{-B}[R_y - (p+1)(\Delta \Omega + \frac{p+2}{4}(\partial \Omega)^2)]\,.
\end{equation}

If we consider that the dilaton field can be decomposed in a compact
and a non-compact dependence: $\phi(x,y) = \varphi(x) + \phi(y)$. The
action for the dilaton in the Einstein frame is of the form:
\begin{equation}
S = \frac{-1}{2 \kappa^2} \int \sqrt{-G} \frac{1}{2}(e^{-\Omega}(\partial
\varphi(x))^2 + e^{-B}(\partial \phi(y))^2)\,.
\end{equation}
This term  has a contribution to the energy momentum tensor:
\begin{eqnarray}
T^{10}_{\mu \nu} & = & \frac{1}{2}
\partial_{\mu}\varphi(x)\partial_{\nu}\varphi(x) - \frac{1}{4} e^\Omega g_{\mu
\nu} ( e^{-\Omega}(\partial \varphi(x))^2 + e^{-B}(\partial \phi(y))^2)\,,
\nonumber 
\\ T^{10}_{\mu i} & = & \frac{1}{2}
\partial_{\mu}\varphi(x)\partial_{i}\phi(y)\,, \nonumber 
\\ T^{10}_{ij} & =
& \frac{1}{2} \partial_{i}\phi(y)\partial_{j}\phi(y) - \frac{1}{4} e^B
g_{ij} ( e^{-\Omega}(\partial \varphi(x))^2 + e^{-B}(\partial \phi(y))^2)\,.
\end{eqnarray}


\end{document}